\documentclass[a4paper,twocolumn,10pt]{article}
\usepackage{free-format}

\newcommand{\KA}{\mathcal{K}_{A}}
\newcommand{\KB}{\mathcal{K}_{B}}
\newcommand{\Enc}{\mathsf{Enc}}
\newcommand{\Dec}{\mathsf{Dec}}
\newcommand{\pk}{\mathsf{pk}}
\newcommand{\sk}{\mathsf{sk}}
\newcommand{\R}{\mathbb{R}}

\runningtitle{Koopman Lifting for Linear Encrypted Control of Nonlinear Systems}
\title{
Koopman Lifting for Linear Encrypted Control of Nonlinear Systems: \\
Applications to LQR and MPC}

\authorlist{ \corrauthor{Uta Kawakami}{1}{0009-0002-8631-8388}\quad \aut{Kenji Sawada}{1}{0000-0001-8935-0434} }

\affiliationlist{ \aff{1}{Graduate School of Engineering, The University of Osaka, Suita, Osaka 565-0871, Japan} }

\correspondingemail{uta.kawakami@misc.mech.eng.osaka-u.ac.jp}
\otheremails{Kenji Sawada: knj.sawada@mech.eng.osaka-u.ac.jp}

\begin{document}
\maketitle
\frontnotes{%
\statement{Funding Statement}{This work was supported by JST BOOST under Grant No. JPMJBS2402; JST CREST under Grant Nos. JPMJCR23M4 and JPMJCR23M2.}
\statement{Conflict of Interest Statement}{The authors declare no conflicts of interest.}
\statement{Data Availability Statement}{The data that support the findings of this study are available from the corresponding author upon reasonable request.}
\statement{Generative AI Statement}{ChatGPT (OpenAI) and Claude (Anthropic) were used for translation and language editing. The authors are responsible for the content.}
\statement{Word Count}{Approximately 5,300 words, including references.}
}

\begin{abstract}
    Homomorphic encryption enables control computations to be outsourced to a third-party cloud server without disclosing measurements or controller parameters.
    Applying existing encrypted linear controllers to nonlinear plants requires a compatible representation of the plant dynamics.
    This paper presents Homomorphically Encrypted Koopman-Based Control (HEK-Control), which connects nonlinear plants to encrypted linear computations through Koopman lifting.
    A trusted plant-side client evaluates nonlinear functions of the measured state, and a linear predictor identified in these coordinates supports conventional controller design.
    The resulting control laws retain the server's encrypted linear operations.
    We demonstrate the framework using state feedback based on linear quadratic regulation and input-constrained model predictive control.
    Closed-loop simulations with the Cheon--Kim--Kim--Song encryption scheme show close agreement with unencrypted computations.
    Additional unencrypted simulations compare regulation under input constraints across a common set of controller designs.
    For a continuous stirred-tank reactor, Koopman lifting supports regulation at input bounds not handled by the tested controllers designed directly in the original state coordinates.
    Tests of encrypted computations also show that packing multiple values into each ciphertext reduces computation time and communication volume.
    These results clarify how existing encrypted linear computations can serve nonlinear plants.
\end{abstract}


\section{Introduction}\label{sec:intro}
    Cloud computing provides a means of outsourcing control computations to servers operated by a third party.
    In this client--server architecture, a plant-side client sends measurements to a cloud server that computes the control inputs.
    This arrangement enables resource sharing and centralized controller management~\cite{darup2021}.
    However, processing measurements and controller parameters in plaintext may expose plant operating conditions and control policies.
    Transport Layer Security (TLS) protects these data in transit between the client and server~\cite{tls13}.
    Since the server decrypts the received data for conventional computation, channel encryption alone does not conceal these data from the server during processing.
    Encrypted control addresses this need to preserve confidentiality during server-side computation.

    Encrypted control meets this requirement by using homomorphic encryption (HE) to perform the supported control operations on ciphertexts~\cite{kogiso2015}.
    For example, the Cheon--Kim--Kim--Song (CKKS) scheme supports approximate addition and multiplication of real-valued data~\cite{cheon2017}.
    Trigonometric functions and comparisons are not directly available, and deeper chains of ciphertext multiplications require larger cryptographic parameters and more computation.
    The delegated control law must therefore be represented using few homomorphic additions and multiplications.

    The compatibility of linear control laws with these arithmetic restrictions has motivated extensive work on encrypted linear controllers~\cite{darup2021}.
    Encrypted state-feedback implementations protect state measurements and feedback gains during outsourced evaluation~\cite{kogiso2015,tran2020}.
    For model predictive control (MPC), client-assisted implementations evaluate affine optimization steps at the server and constraint projections at a trusted client~\cite{alexandru2018}.
    Extending these implementations to nonlinear plants requires a plant representation compatible with the existing encrypted linear computations.

    Several approaches have been investigated to handle nonlinearity in encrypted control of nonlinear plants.
    Encrypted linear feedback has been used to stabilize an inverted pendulum near an equilibrium~\cite{tran2020}.
    For an antagonistic pneumatic artificial-muscle actuator, a nonlinear control law is approximated by polynomials.
    The client evaluates nonlinear features and delegates an encrypted inner product to the server~\cite{takeda2025}.
    In piecewise-affine control, the client identifies the region containing the state, and the server evaluates the corresponding affine law~\cite{darup2018explicit}.
    Encrypted feedback linearization similarly evaluates model-dependent nonlinear terms at the client and delegates their linear combination~\cite{teranishi2020}.
    These approaches simplify server-side computation, while their applicability and controller representations depend on local models, region selection, or plant-specific transformations.

    A common design procedure is therefore needed to reuse encrypted linear-control implementations across nonlinear plants.
    Local linearization is applicable in a neighborhood of the expansion point, whereas approximating a given nonlinear control law requires an arithmetic representation tailored to that law.
    A linear predictor identified from data over the operating region would instead allow established linear controller designs to be used without a plant-specific analytical linearization.
    For an architecture that evaluates nonlinear functions at a trusted client while retaining server-side encrypted operations, control performance and implementation costs must be assessed using an established HE scheme at a standard security level.

    This study proposes Homomorphically Encrypted Koopman-Based Control (HEK-Control), which uses Koopman lifting to extend existing encrypted linear-control implementations to nonlinear plants.
    We evaluate closed-loop performance under input constraints and the associated computation and communication costs.
    The client evaluates nonlinear observables, and a controller is designed offline for a linear predictor identified in these coordinates~\cite{korda2018}.
    This construction retains the server-side encrypted linear operations.
    State feedback designed by linear quadratic regulation (LQR) and MPC serve as representative applications.

    Unlike approximating a prescribed nonlinear control law, HEK-Control identifies a predictor of the plant and uses it to design the controller in observable coordinates.
    The contribution lies in connecting this data-driven plant representation to existing encrypted control implementations and evaluating the resulting closed loop, rather than in a new encryption scheme or optimization algorithm.

    The main contributions of this study are as follows.
    \begin{enumerate}[label=(C\arabic*),leftmargin=*,itemsep=2pt,topsep=3pt]
        \item We construct HEK-Control by evaluating Koopman lifting at the client, enabling nonlinear plants to be controlled while retaining server-side encrypted linear operations.
        \item We apply the framework to encrypted state feedback and encrypted MPC on nonlinear benchmarks, retaining their controller-design methods and constraint handling.
        \item We implement HEK-Control using CKKS configured for 128-bit classical security and quantify numerical agreement, control performance, and computation and communication costs, including the effect of ciphertext packing.
    \end{enumerate}

    This paper extends the position paper presented at SICE FES 2026~\cite{kawakami2026}.
    That study demonstrated a proof of concept by applying fixed-gain LQR to three benchmarks with different nonlinearities using numerical emulation of CKKS arithmetic.
    The present work evaluates the closed loop using actual CKKS encryption, homomorphic evaluation, and decryption.
    The evaluation uses five different nonlinear benchmarks and covers setpoint tracking, iterative input-constrained MPC, and computational and communication costs.

    Section~\ref{sec:prelim} introduces the preliminaries, and Section~\ref{sec:problem} defines the problem.
    Section~\ref{sec:method} presents HEK-Control and its connections to the two control methods.
    Sections~\ref{sec:setup} and~\ref{sec:results} describe the experimental conditions and results.
    Section~\ref{sec:discussion} discusses the scope and limitations, and Section~\ref{sec:conclusion} concludes the paper.

\section{Preliminaries}\label{sec:prelim}
    This section introduces the homomorphic encryption, encrypted linear control, and Koopman operator theory underlying the proposed method.

    \subsection{Homomorphic Encryption}
        For the CKKS scheme, let $\Enc_{\pk}$ denote encoding followed by encryption with the public key $\pk$, and let $\Dec_{\sk}$ denote decryption with the secret key $\sk$ followed by decoding.
        For ciphertexts encrypted under the same key with compatible levels and scales, homomorphic addition $\oplus$ and multiplication $\otimes$ satisfy
        \begin{equation}
            \begin{aligned}
                \Dec_{\sk}\!\left(\Enc_{\pk}(a)\oplus\Enc_{\pk}(b)\right)
                &\simeq a+b,\\
                \Dec_{\sk}\!\left(\Enc_{\pk}(a)\otimes\Enc_{\pk}(b)\right)
                &\simeq a\odot b.
            \end{aligned}\label{eq:he_operations}
        \end{equation}
        Here, $a,b$ are real vectors of the same dimension, $\odot$ denotes componentwise multiplication, and $\simeq$ denotes approximate equality including encoding and homomorphic arithmetic errors.
        We omit key subscripts below and initially express the operations with each scalar entry encrypted separately.

        Multiplicative depth counts ciphertext multiplications along the longest dependency path.
        Deeper CKKS computations require a larger modulus to retain precision as rescaling consumes ciphertext levels.
        Maintaining the security level may then require a larger ring dimension, increasing computation time and ciphertext size~\cite{cheon2017,suh2021}.

    \subsection{Encrypted Linear Control}\label{sec:linearcontrol}
        For $M\in\R^{p\times q}$ and $v\in\R^q$, encrypting both operands gives
        \begin{equation}
            \Dec\!\left(\bigoplus_{j=1}^{q}
            \Enc(M_{ij})\otimes\Enc(v_j)\right)
            \simeq (Mv)_i,
            \label{eq:encrypted_matvec}
        \end{equation}
        by combining the CKKS operations in~\eqref{eq:he_operations}~\cite{cheon2017}.
        Evaluating all $p$ outputs in this way requires $pq$ ciphertext multiplications at depth one.

        \subsubsection{Encrypted LQR}\label{sec:prelim_lqr}
            For the linear prediction model
            \begin{equation}
                \chi_{k+1}=A\chi_k+Bv_k,
                \qquad \chi_k\in\R^d,\quad v_k\in\R^m,
                \label{eq:linear_model}
            \end{equation}
            consider minimizing the finite-horizon quadratic objective
            \begin{equation}
                J_H=\sum_{j=0}^{H-1}\!\left(
                \chi_{j|k}^{\top}Q\chi_{j|k}+v_{j|k}^{\top}Rv_{j|k}
                \right)+\chi_{H|k}^{\top}P_{\mathrm f}\chi_{H|k},
                \label{eq:objective}
            \end{equation}
            with $Q\succeq0$, $R\succ0$, and terminal weight $P_{\mathrm f}\succeq0$.
            The index $j|k$ denotes a $j$-step prediction at time $k$, with $\chi_{0|k}=\chi_k$.
            Without input constraints, the backward Riccati recursion from $P_H=P_{\mathrm f}$,
            \begin{equation}
                \begin{aligned}
                    K_j&=(R+B^\top P_{j+1}B)^{-1}B^\top P_{j+1}A,\\
                    P_j&=Q+A^\top P_{j+1}(A-BK_j),
                \end{aligned}\label{eq:riccati}
            \end{equation}
            for $j=H-1,\ldots,0$ gives the first input $v_k=-K_0\chi_k$~\cite{anderson1990}.
            Here, LQR denotes fixed-gain state feedback using $K=K_0$ from this finite-horizon design with $P_{\mathrm f}=Q$.
            The encrypted control law is evaluated by substituting $M=-K$ and $v=\chi_k$ into~\eqref{eq:encrypted_matvec}.

        \subsubsection{Encrypted MPC}\label{sec:prelim_mpc}
            MPC adds the input constraint $v_{j|k}\in\mathcal V$ to the minimization of~\eqref{eq:objective} and solves the problem at each sampling instant.
            Here, $\mathcal V$ is a closed convex input set.
            Eliminating the predicted states gives the quadratic program (QP) in the stacked input sequence $U\in\R^{Hm}$,
            \begin{equation}
                \min_{U\in\mathcal V^H}\;
                \tfrac12 U^\top\mathcal H U+U^\top\mathcal F\chi_k,
                \label{eq:qp}
            \end{equation}
            where $\mathcal H\succ0$ and $\mathcal F$ are computed offline from the prediction model and weights~\cite{rawlings2017}.

            The client-assisted implementation uses $L$ projected fast-gradient iterations~\cite{alexandru2018}.
            Starting from $U^0=Y^0=0$, it computes
            \begin{equation}
                \begin{aligned}
                    U^{\ell+1}&=\Pi_{\mathcal V^H}\!\left(
                    (I-\eta\mathcal H)Y^\ell-\eta\mathcal F\chi_k\right),\\
                    Y^{\ell+1}&=U^{\ell+1}+\beta(U^{\ell+1}-U^\ell),\\
                    &\hspace{20mm}\ell=0,\ldots,L-1.
                \end{aligned}\label{eq:fgm}
            \end{equation}
            The Euclidean projection $\Pi_{\mathcal V^H}$ acts separately on each predicted input.
            The iteration parameters are $\eta=1/\lambda_{\max}(\mathcal H)$, $\kappa=\lambda_{\max}(\mathcal H)/\lambda_{\min}(\mathcal H)$, and $\beta=(\sqrt\kappa-1)/(\sqrt\kappa+1)$, where $\lambda_{\max}$ and $\lambda_{\min}$ denote the largest and smallest eigenvalues of $\mathcal H$.
            After $L$ iterations, the first $m$ entries of the input sequence are applied, and the optimization is repeated at the next sampling instant.

    \subsection{Koopman Operator Theory}\label{sec:koopman}
        For the discrete-time state-transition map $F(x,u)$, fix an input $\bar u$ and define $F_{\bar u}(x)=F(x,\bar u)$.
        A real-valued function $g$ of the state is called an observable.
        The Koopman operator $\mathcal K_{\bar u}$ associated with $F_{\bar u}$ acts on such functions according to
        \begin{equation}
            (\mathcal K_{\bar u}g)(x)=g(F_{\bar u}(x)),
            \label{eq:koopman_operator}
        \end{equation}
        so that the next value of $g$ is expressed as a function of the current state~\cite{koopman1931,bevanda2021}.
        Regardless of the nonlinearity of $F_{\bar u}$, this operator satisfies
        \begin{equation}
            \mathcal K_{\bar u}(\alpha_1g_1+\alpha_2g_2)
            =\alpha_1\mathcal K_{\bar u}g_1+\alpha_2\mathcal K_{\bar u}g_2,
            \label{eq:koopman_linearity}
        \end{equation}
        for any real scalars $\alpha_1,\alpha_2$.

        For finite-dimensional computation, choose a dictionary $\psi=[g_1,\ldots,g_N]^\top$ and define $z_k=\psi(x_k)$.
        If each next-step observable is a linear combination of this dictionary, its span is invariant under the operator and admits an exact matrix representation~\cite{brunton2016}.
        Otherwise, these combinations approximate the next-step values~\cite{bevanda2021}.
        Extended dynamic mode decomposition (EDMD) finds their coefficients by least-squares fitting to observable transitions~\cite{williams2015}.

        For time-varying control inputs, identification also includes the input data.
        Extended dynamic mode decomposition with control (EDMDc) constructs the linear predictor
        \begin{equation}
            z_{k+1}\approx\KA z_k+\KB u_k,
            \label{eq:predictor}
        \end{equation}
        from these data~\cite{korda2018}.
        For input dimension $m$, the matrices satisfy $\KA\in\R^{N\times N}$ and $\KB\in\R^{N\times m}$.
        For $S$ lifted transitions $(z^{(r)},u^{(r)},z^{(r),+})$, let $A$ and $B$ denote candidate matrices with the same dimensions as $\KA$ and $\KB$.
        Their optimal values are obtained from
        \begin{equation}
            \begin{aligned}
                (\KA,\KB)=\arg\min_{A,B}\;&
                \sum_{r=1}^{S}\!\left\|z^{(r),+}
                -Az^{(r)}-Bu^{(r)}\right\|_2^2\\
                &+\lambda\|[A\;B]\|_F^2.
            \end{aligned}\label{eq:edmdc}
        \end{equation}
        Here, $z^{(r),+}$ is the observable vector evaluated at the successor state, $\lambda>0$ is the regularization coefficient, and $\|\cdot\|_F$ is the Frobenius norm.

\section{Problem Setting}\label{sec:problem}
    This section specifies the plant, control objectives, and confidentiality assumptions for outsourced control.
    \subsection{Control Problem}
        Consider a nonlinear plant whose full state is measured at sampling period $T_s$:
        \begin{equation}
            x_{k+1}=F(x_k,u_k),\qquad x_k\in\R^n,\quad u_k\in\R^m.
            \label{eq:plant}
        \end{equation}
        Regulation targets an equilibrium $(x^\star,u^\star)$, while tracking switches between reference equilibria with admissible steady-state inputs.
        We use the normalized deviation coordinates
        \begin{equation}
            \xi_k=D_x^{-1}(x_k-x^\star),\qquad
            \nu_k=D_u^{-1}(u_k-u^\star),
            \label{eq:normal}
        \end{equation}
        where $D_x$ and $D_u$ are positive diagonal matrices containing representative scales of the states and inputs.
        Normalization places variables with different units and magnitudes on comparable scales for observable evaluation, model identification, and encoding with a common CKKS scale~\cite{cheon2017,higham2002}.

        Let $\bar\nu$ collect the normalized input bounds and define $\mathcal U=\{\nu:|\nu_i|\leq\bar\nu_i\}$.
        Here, $\bar\nu=D_u^{-1}u_{\max}$, where $u_{\max}$ bounds the physical input deviation from $u^\star$ and is distinct from the normalization scale $D_u$.
        The client limits the decrypted command $\widehat\nu_k$ and applies
        \begin{equation}
            u_k=u^\star+D_u\min\{\bar\nu,\max\{-\bar\nu,\widehat\nu_k\}\},
            \label{eq:applied}
        \end{equation}
        where the minimum and maximum are taken componentwise.

    \subsection{Threat Model}
        We consider a client--server control architecture in which the plant-side client acquires measurements and applies control inputs while outsourcing control computations to a third-party cloud server~\cite{darup2021,alexandru2018}.
        The plant and client are trusted, whereas the server follows the prescribed computations but may attempt to infer confidential information from received data (honest-but-curious).
        The client performs identification, controller design, key management, and all encryption.
        The server receives encrypted observables and coefficients, cryptographic parameters, and evaluation keys for homomorphic computation.
        The secret key and decrypted values remain with the client and are not returned to the server~\cite{openfhesecurity}.
        Vector dimensions, operation order, and iteration counts are public.

        We protect the confidentiality of measurements, observables, coefficients, and control commands.
        Client compromise, direct observation of plant signals, and active server attacks are excluded.
        Integrity, authentication, replay prevention, and availability require separate measures~\cite{teranishi2022security}.

\section{Proposed Method}\label{sec:method}
\begin{figure*}[t]
        \centering
        \captionsetup{font=footnotesize}
        \captionsetup[subfigure]{font=footnotesize}
        \includegraphics[width=0.83\textwidth]{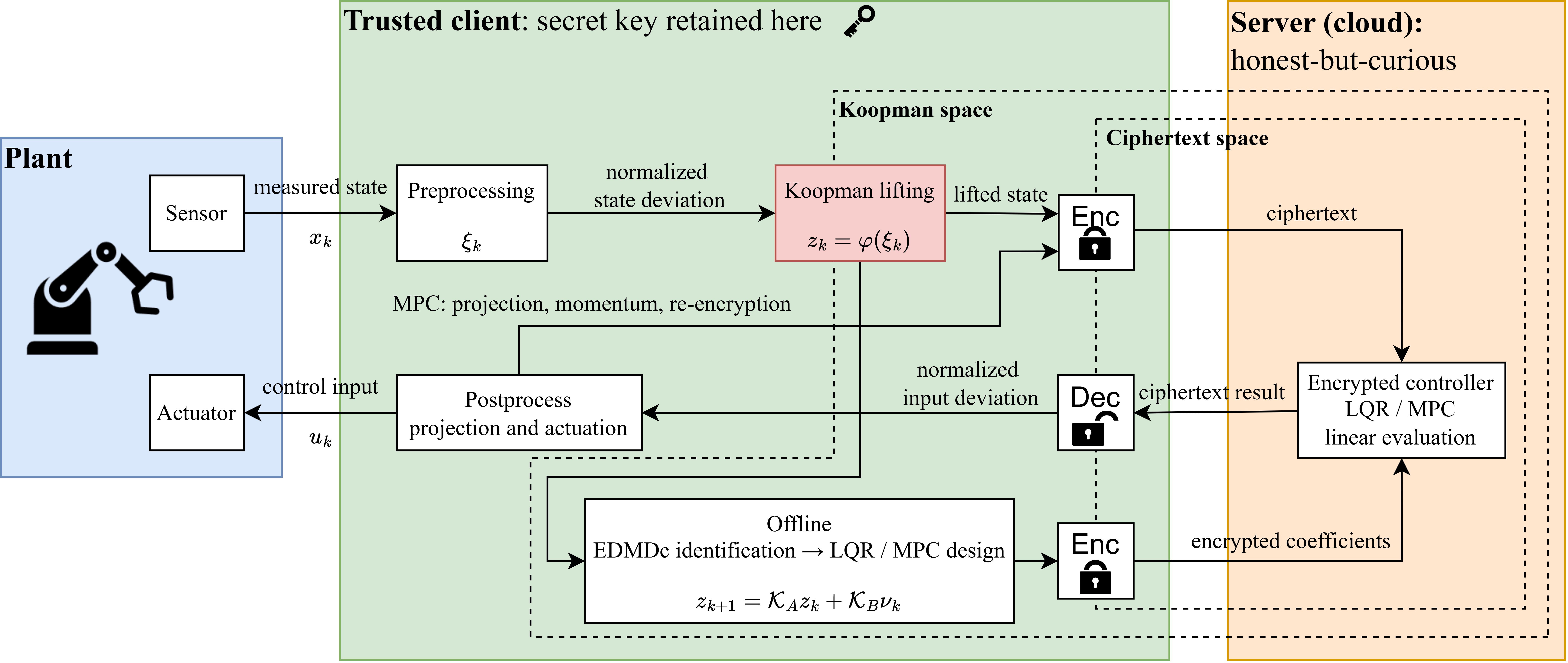}
        \caption{HEK-Control architecture. The trusted client performs offline identification and controller design. The third-party server evaluates encrypted linear operations using encrypted observables and coefficients.}
        \label{fig:architecture}
    \end{figure*}

    HEK-Control connects client-side Koopman lifting to existing encrypted linear control.
    Figure~\ref{fig:architecture} shows the client--server data flow and separates offline controller preparation from online execution.
    Offline, the client identifies the predictor, designs the controller, and encrypts its coefficients.
    Online, the client evaluates and encrypts the observables, and the server computes and returns the encrypted linear transformation.
    After decryption, the client performs the local controller operations and enforces the input limits.
    The following subsections describe predictor identification and the LQR and MPC applications.

    \subsection{Predictor Identification}\label{sec:liftedmodel}
        Choose a dictionary $\psi$ whose first $n$ entries are the normalized states, and define the centered map
        \begin{equation}
            \begin{gathered}
                z_k=\varphi(\xi_k)=\psi(\xi_k)-\psi(0)\in\R^N,\\
                \varphi(0)=0,\qquad \xi_k=Cz_k,\quad C=[I_n\;0].
            \end{gathered}\label{eq:lift}
        \end{equation}
        This lifted state is zero at the target and retains the physical-state coordinates for controller design.

        Applying EDMDc in~\eqref{eq:edmdc} to the transformed transitions and normalized inputs gives $z_{k+1}\approx\KA z_k+\KB\nu_k$.
        This predictor is used for controller design.
        Setting $\varphi(\xi)=\xi$ gives dynamic mode decomposition with control (DMDc), which identifies a linear model using state coordinates alone and serves as a baseline~\cite{proctor2016}.

    \subsection{LQR Application}\label{sec:lqr}
        The LQR design in Section~\ref{sec:prelim_lqr} is applied with $A=\KA$, $B=\KB$, and $\chi_k=z_k$.
        The state weighting is chosen as
        \begin{equation}
            Q_z=C^\top Q_xC,\qquad
            z^\top Q_zz=\xi^\top Q_x\xi,
            \label{eq:lifted_cost}
        \end{equation}
        as in~\cite{korda2018}.
        This preserves the cost on the normalized physical state without an independent penalty on the additional observables $z_{n+1},\ldots,z_N$.
        The fixed gain $K$ from~\eqref{eq:riccati} is evaluated using $M=-K$ and $v=z_k$ in~\eqref{eq:encrypted_matvec}.
        The client applies the decrypted command through~\eqref{eq:applied}.
        The law is linear in $z$ but generally nonlinear in the original state through $\varphi$.

        For setpoint tracking, express the reference equilibrium as $(\xi_{r,k},\nu_{r,k})$ in the same normalized coordinates and use
        \begin{equation}
            \widehat\nu_k\simeq\nu_{r,k}
            -K\bigl(\varphi(\xi_k)-\varphi(\xi_{r,k})\bigr).
            \label{eq:tracking}
        \end{equation}
        The client encrypts the observable difference and adds the steady-state input after decrypting the feedback term.
        The gain and dictionary stay fixed, and~\eqref{eq:applied} limits the total command.

    \subsection{MPC Application}\label{sec:mpc}
        The MPC formulation in Section~\ref{sec:prelim_mpc} is applied with $A=\KA$, $B=\KB$, $Q=Q_z$, $\chi_k=z_k$, and $\mathcal V=\mathcal U$.
        The terminal weight can be chosen as $Q_z$ or as the Riccati matrix $P_{\mathrm{LQR}}$ used to compute the LQR gain for the same predictor.
        LQR and MPC horizons are specified separately in Section~\ref{sec:settings}.
        The matrices $\mathcal F\in\R^{Hm\times N}$ and $\mathcal H\in\R^{Hm\times Hm}$ are recomputed for the predictor and terminal weight.
        Input constraints and the encrypted computation procedure are unchanged.

        The server evaluates $-\eta\mathcal Fz_k$ once and $(I-\eta\mathcal H)Y^\ell$ at each iteration using~\eqref{eq:encrypted_matvec}.
        The client decrypts and adds these terms, applies the projection and momentum update in~\eqref{eq:fgm}, and re-encrypts the iterate.
        Unlike the public-model formulation of the client-assisted method~\cite{alexandru2018}, this implementation also encrypts the coefficient matrices using CKKS.
        Each server call has depth one, and each update requires $L+1$ round trips.

        Table~\ref{tab:counts} summarizes the ciphertext multiplications, transmitted ciphertexts, and communication round trips per update when each scalar is encrypted separately.
        Koopman lifting enlarges the initial state-dependent product, but not the MPC iterate dimension.
        The unconstrained QP and Riccati recursion give the same first move under identical design conditions.
        We evaluate MPC with input constraints.

        \begin{table}[t]
            \centering\footnotesize
            \captionsetup{font=footnotesize}
            \caption{Per-update costs with one ciphertext per scalar. Both operands are encrypted. Initialization is excluded.}
            \label{tab:counts}
            \setlength{\tabcolsep}{3pt}
            \begin{tabularx}{\columnwidth}{@{}Xcc@{}}\toprule
                Quantity & LQR & MPC\\\midrule
                Ciphertext multiplications & $mN$ & $HmN+L(Hm)^2$\\
                Uploaded ciphertexts & $N$ & $N+LHm$\\
                Returned ciphertexts & $m$ & $(L+1)Hm$\\
                Exchanges & $1$ & $L+1$\\
                Depth per server call & $1$ & $1$\\\bottomrule
            \end{tabularx}
        \end{table}

\section{Experimental Setup}\label{sec:setup}
    We evaluate numerical accuracy, control performance, and implementation costs.
    Table~\ref{tab:plants} lists the five plants, observable dictionaries, and tracking references, including an exothermic continuous stirred-tank reactor (CSTR).
    The slow-manifold model and dictionary follow~\cite{brunton2016}.
    The DC motor and its 100 thin-plate radial basis functions (RBFs) follow~\cite{korda2018}, with full-state measurements and symmetric input bounds about the nominal input.

    \begin{table*}[t]
        \centering\footnotesize
        \captionsetup{font=footnotesize}
        \caption{Plants, observable dictionaries, and tracking references ($0,+a,0,-a,0$).}
        \label{tab:plants}
        \begin{tabularx}{\textwidth}{@{}lccXX@{}}\toprule
            Plant & $n/m$ & $N$ & Observable dictionary & Reference amplitude / dwell\\\midrule
            CSTR & $2/1$ & 5 & Nonconstant monomials of degree at most 2 & Temperature: $\pm5\,\mathrm K$ / $2.5\,\mathrm{min}$\\
            Slow manifold & $2/1$ & 3 & $\xi_1,\xi_2,\xi_1^2$ & Second state: $\pm2$ / $45\,\mathrm s$\\
            van der Pol & $2/1$ & 20 & Nonconstant monomials of degree at most 5 & First state: $\pm0.8$ / $10\,\mathrm s$\\
            Reaction--diffusion & $16/3$ & 63 & States, squares, cubes, adjacent-state products & Midpoint field: $\pm0.2$ / $2.5\,\mathrm s$\\
            DC motor & $2/1$ & 102 & States and 100 thin-plate radial basis functions & Speed: $\pm30\,\mathrm{rad/s}$ / $0.5\,\mathrm s$\\\bottomrule
        \end{tabularx}
    \end{table*}

    \subsection{Comparison Models}
        The baselines use a Jacobian linearization at the nominal equilibrium or a DMDc model identified from state data.
        DMDc and EDMDc share normalized data and regression, differing only in the observable map.
        The Jacobian uses central differences of the sampled plant map with step $10^{-6}$.
        All methods use the same weights, initial states, and input bounds.
        Reference equilibria and feedforward inputs are also common to all methods and are computed from the plant's steady-state equations.

        Table~\ref{tab:settings} specifies the sampling periods, data sizes, control weights, normalization scales, and initial states.
        The training box is $[-1.2,1.2]^n$, except for van der Pol, where it is $[-1.6,1.6]\times[-1.4,1.4]$.
        Initial states for identification and the motor RBF centers are sampled uniformly from this box.
        Trajectories use zero-mean Gaussian input excitation clipped to the input bounds.
        Data and RBF-center seeds are 42 and 7, respectively.
        The ridge coefficient is $10^{-6}$ times the mean diagonal of the regression Gram matrix.

        \begin{table*}[t]
            \centering\footnotesize
            \captionsetup{font=footnotesize}
            \caption{Identification and control settings. $S$: trajectories $\times$ steps. $s_i$: spatial position.}
            \label{tab:settings}
            \setlength{\tabcolsep}{4pt}
            \begin{tabularx}{\textwidth}{@{}lcccccX@{}}\toprule
                Plant & $T_s$ & $S$ & $(q_x,r)$ & $\operatorname{diag}(D_x)$ & $\operatorname{diag}(D_u)$ & Initial state $x_0$\\\midrule
                CSTR & $0.01\,\mathrm{min}$ & $400\times60$ & $(20,0.1)$ & $(0.15,10)$ & $15$ & $x^\star+(0.18,-12)^\top$\\
                Slow manifold & $0.05\,\mathrm s$ & $200\times40$ & $(1,1)$ & $(5,5)$ & $5$ & $(-5,5)^\top$\\
                van der Pol & $0.02\,\mathrm s$ & $200\times40$ & $(10,0.2)$ & $(2,4)$ & $2$ & $(2,0)^\top$\\
                Reaction--diffusion & $0.02\,\mathrm s$ & $400\times40$ & $(10,0.1)$ & $0.5$ (all) & $2$ (all) & $x_{0,i}=0.55\sin(\pi s_i)$\\
                DC motor & $0.01\,\mathrm s$ & $200\times40$ & $(10,0.1)$ & $(5,60)$ & $2$ & $(0,60)^\top$\\\bottomrule
            \end{tabularx}
        \end{table*}

    \subsection{Implementation}\label{sec:settings}
        Encrypted LQR comparisons use the common horizon $H=31$, $Q_x=q_xI_n$, $R=rI_m$, and terminal weight $Q_z$.
        This horizon provides a common comparison setting and is not assumed to be optimal for each predictor.
        Starting from $Q_z$, 30 Riccati updates give $P_{\mathrm{LQR}}$ for gain computation and MPC terminal weighting.
        Encrypted CSTR MPC uses $H=5$, $L=15$, and $P_{\mathrm f}=P_{\mathrm{LQR}}$.
        MPC and saturated LQR are evaluated for 4 min at 1.0 and 0.6 times the nominal input bound.
        The momentum calculation floors the smallest eigenvalue at $10^{-9}\lambda_{\max}(\mathcal H)$.

        OpenFHE 1.5.1~\cite{openfhe2022} runs on an Intel Core i7-13700F CPU with eight OpenMP threads.
        Client and server share one process and cryptographic context, with no key transfer or network transport.
        Table~\ref{tab:crypto} lists the CKKS parameters used in the closed-loop simulations, which store each scalar in a separate ciphertext.
        Coefficients are encrypted once, and every transmitted observable vector or optimization iterate is encrypted anew.
        Values below $2^{-32}$ in absolute value are zeroed before encoding.
        Plant dynamics are integrated by the fourth-order Runge--Kutta method with zero-order-held inputs.
        Measurement noise and computation and communication delays are omitted from the closed-loop simulations.

        \begin{table}[t]
            \centering\footnotesize
            \captionsetup{font=footnotesize}
            \caption{CKKS settings. SIMD retains the security level, ring dimension, depth, and scale. The last two rows apply without SIMD.}
            \label{tab:crypto}
            \begin{tabularx}{\columnwidth}{@{}lX@{}}\toprule
                Item & Setting\\\midrule
                Security & \texttt{HEStd\_128\_classic}\\
                Ring dimension & 8192\\
                Depth & 1 per server call\\
                Scale / first modulus & $2^{40}$ / 50 bits\\
                Scaling technique & \texttt{FLEXIBLEAUTO}\\
                Slots used & 1 (minimum batch size 8)\\
                Reported ciphertext size & 256.8 KiB\\\bottomrule
            \end{tabularx}
        \end{table}

        The effect of single-instruction multiple-data (SIMD) packing is measured separately on the same platform, using encrypted random matrices and vectors with the same dimensions as the control computations.
        For $M\in\R^{p\times q}$, the vector $v$ is padded to $b=2^{\lceil\log_2q\rceil}$ entries and replicated in $p$ blocks aligned with the matrix rows.
        When $pb\leq4096$, one ciphertext holds the blocks.
        One multiplication and $\log_2b$ rotations and additions yield the outputs at block starts at depth one, with additional rotation keys~\cite{cheon2017}.
        Reported times are medians of 30 runs with eight OpenMP threads and include encryption, homomorphic evaluation, and decryption.
        Initialization and network transport are excluded.

    \subsection{Evaluation Metrics}
        The performance index uses actual states and applied inputs:
        \begin{equation}
            \begin{aligned}
                J=T_s\sum_k\bigl[&(\xi_k-\xi_{r,k})^\top Q_x(\xi_k-\xi_{r,k})\\
                &+(\nu_k-\nu_{r,k})^\top R(\nu_k-\nu_{r,k})\bigr].
            \end{aligned}\label{eq:metric}
        \end{equation}
        Regulation has zero reference, and tracking includes the initial regulation interval.
        Numerical error is the maximum absolute normalized-input difference from unencrypted LQR with the same gain or MPC with the same 15 iterations.
        An initial-state grid is a set of uniformly spaced initial conditions in normalized state coordinates.
        The settling rate is the percentage of these conditions for which $\|\xi_k\|_2\leq0.02$ is reached and maintained until the evaluation ends.
        The CSTR MPC comparison uses a common $25\times25$ grid with 8-min unencrypted runs, distinct from the 4-min encrypted trajectories.
        Communication volume is estimated from ciphertext payload counts.

        Tuning sensitivity is evaluated without encryption over twelve designs: LQR with $H\in\{11,16,21,31,46,61\}$ and MPC with $H\in\{5,10,20\}$, $P_{\mathrm f}\in\{Q_z,P_{\mathrm{LQR}}\}$, and $L=15$.
        DMDc and EDMDc share identification data generated separately from the encrypted comparisons under the same conditions.
        Each predictor's maximum settling rate is selected and evaluated on the same $11\times11$ grid over the training region, not on held-out initial states.
        The intervals are 8 min for CSTR and 20 s for van der Pol, with a 95\% settling criterion.
        CSTR data sensitivity uses a $15\times15$ grid and ten seeds (42 and 1--9).
        Van der Pol is also evaluated over twice the training region at the nominal bound.

\section{Experimental Results}\label{sec:results}
    We first check numerical agreement with unencrypted control, then compare LQR and MPC performance.
    Computation time and communication volume are assessed separately.

    \subsection{Numerical Agreement}
        Each encrypted controller is compared with its unencrypted counterpart under the same settings.
        The maximum normalized-input differences are $8.0\times10^{-8}$ for LQR regulation and $2.2\times10^{-7}$ for tracking, with relative cost differences at most $1.1\times10^{-9}$.
        For CSTR MPC with $P_{\mathrm{LQR}}$, the maximum input difference is $4.1\times10^{-7}$.
        These CKKS errors are small compared with the control-performance differences below.

    \subsection{LQR Performance}\label{sec:lqrresults}
        At $H=31$, HEK-Control gives lower costs than both baselines for CSTR, the slow-manifold system, and van der Pol.
        Slow-manifold regulation costs are 217.4, 151.5, and 45.81 for Jacobian, DMDc, and HEK-Control, respectively.
        The corresponding reaction--diffusion costs are 3.193, 3.193, and 3.194, and all three motor costs are 0.4866.

\begin{figure*}[t]
            \centering
            \captionsetup{font=footnotesize}
            \captionsetup[subfigure]{font=footnotesize}
            \includegraphics[width=0.83\textwidth]{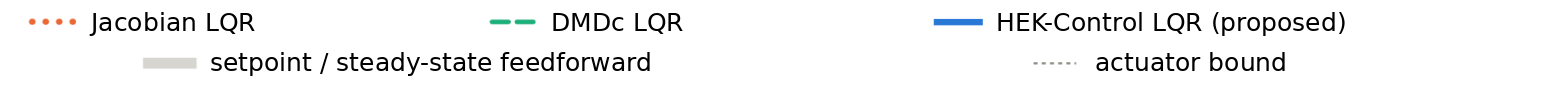}\par
            \begin{minipage}[t]{0.30\textwidth}
                \centering
                \includegraphics[width=\linewidth]{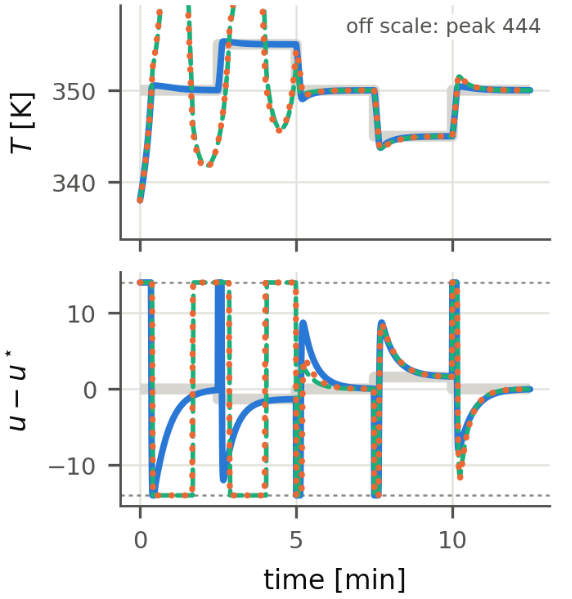}
                \subcaption{CSTR}\label{fig:tracking_cstr}
            \end{minipage}\hfill
            \begin{minipage}[t]{0.30\textwidth}
                \centering
                \includegraphics[width=\linewidth]{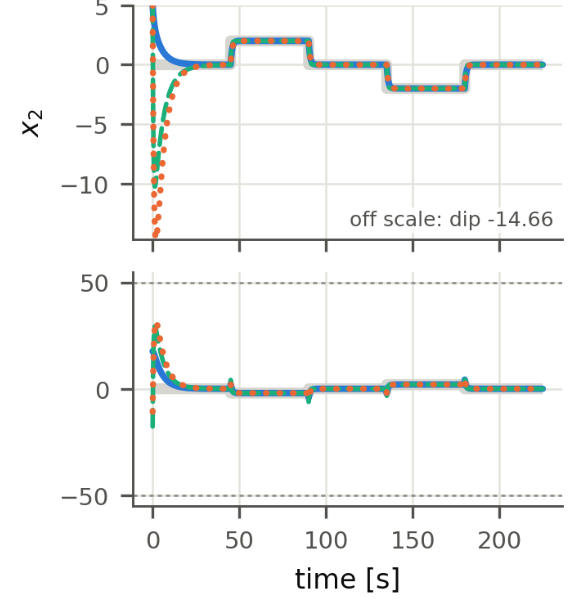}
                \subcaption{Slow-manifold system}\label{fig:tracking_slow}
            \end{minipage}\hfill
            \begin{minipage}[t]{0.30\textwidth}
                \centering
                \includegraphics[width=\linewidth]{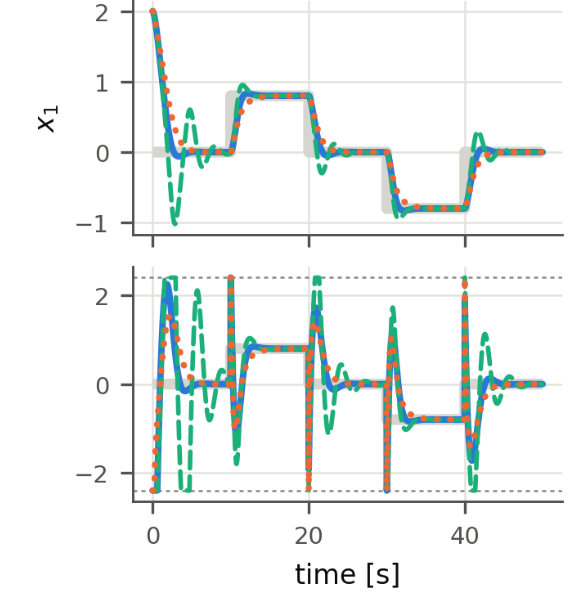}
                \subcaption{Van der Pol oscillator}\label{fig:tracking_vdp}
            \end{minipage}\par
            \caption{Encrypted LQR tracking ($H=31$): output (top) and input (bottom). Annotated extrema are outside the axes. CSTR input deviation and temperature are in kelvin.}
            \label{fig:tracking}
        \end{figure*}

        Figure~\ref{fig:tracking} compares tracking responses at this horizon.
        HEK-Control avoids the large CSTR temperature excursion of both baselines.
        It also reduces the initial undershoot in the slow-manifold system and the transient oscillations of DMDc control for van der Pol.
        Later slow-manifold tracking segments are similar, so its improvement mainly concerns the initial transient.
        At $H=11$, all three models track the CSTR references with comparable costs, showing that Fig.~\ref{fig:tracking}(a) depends on horizon selection.

    \subsection{Constrained MPC}\label{sec:mpcresults}
\begin{figure*}[t]
            \centering
            \captionsetup{font=footnotesize}
            \captionsetup[subfigure]{font=footnotesize}
            \includegraphics[width=0.57\textwidth]{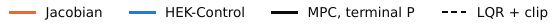}\par
            \begin{minipage}[t]{0.38\textwidth}
                \centering
                \includegraphics[width=\linewidth]{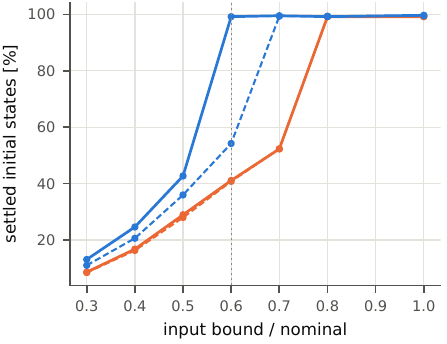}
                \subcaption{Settling success (unencrypted)}\label{fig:mpc_success}
            \end{minipage}\hfill
            \begin{minipage}[t]{0.47\textwidth}
                \centering
                \includegraphics[width=\linewidth]{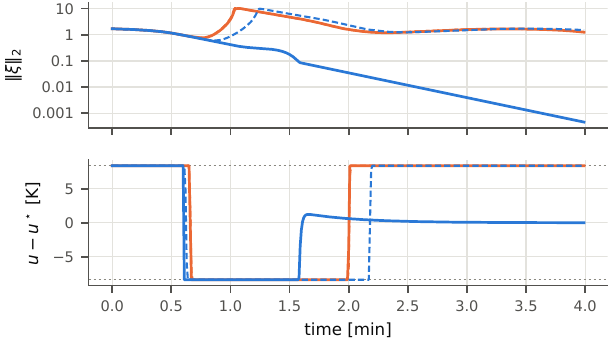}
                \subcaption{Encrypted response at the 60\% input bound}\label{fig:mpc_time}
            \end{minipage}\par
            \caption{CSTR MPC ($H=5$, $L=15$, $P_{\mathrm f}=P_{\mathrm{LQR}}$) versus saturated LQR: (a) 8-min unencrypted grid evaluation. (b) 4-min encrypted response. Legend: \textit{terminal P} is $P_{\mathrm{LQR}}$, and \textit{LQR + clip} is saturated LQR.}
            \label{fig:mpc_response}
        \end{figure*}

        Figure~\ref{fig:mpc_response} compares CSTR settling rates and an encrypted response under input constraints.
        Table~\ref{tab:mpc} gives the corresponding costs, saturation rates, and settling rates for MPC and saturated LQR.
        All controllers settle at the nominal bound.
        At 60\%, only HEK-Control MPC settles from the nominal initial state within the 4-min encrypted run.
        In the separate 8-min unencrypted evaluation, it achieves a 99\% settling rate, compared with 41--54\% for the other controllers.

        \begin{table}[t]
            \centering\footnotesize
            \captionsetup{font=footnotesize}
            \caption{CSTR MPC ($H=5$, $L=15$, $P_{\mathrm f}=P_{\mathrm{LQR}}$) versus saturated LQR. $J$ and saturation (Sat., \%) use 4-min encrypted runs. Grid (\%) is 8-min unencrypted settling. $\dagger$: nominal initial state does not settle.}
            \label{tab:mpc}
            \setlength{\tabcolsep}{3pt}
            \begin{tabularx}{\columnwidth}{@{}cXrrr@{}}\toprule
                Bound & Predictor / control & $J$ & Sat. & Grid\\\midrule
                $1.0$ & Jacobian / MPC & 591.7 & 67.2 & 99\\
                $1.0$ & Jacobian / LQR & 592.7 & 67.0 & 99\\
                $1.0$ & HEK-Control / MPC & 19.65 & 8.7 & 100\\
                $1.0$ & HEK-Control / LQR & 19.08 & 10.5 & 100\\
                $0.6$ & Jacobian / MPC & $782.7^{\dagger}$ & 99.5 & 41\\
                $0.6$ & Jacobian / LQR & $788.0^{\dagger}$ & 99.0 & 41\\
                $0.6$ & HEK-Control / MPC & \textbf{29.87} & 39.5 & \textbf{99}\\
                $0.6$ & HEK-Control / LQR & $711.4^{\dagger}$ & 99.0 & 54\\\bottomrule
            \end{tabularx}
        \end{table}

        Increasing $L$ from 15 to 100 in unencrypted CSTR tests did not change whether the trajectories settled.

        Table~\ref{tab:scope} reports the highest settling rate for each predictor over twelve designs using separately generated identification data.
        Unlike Fig.~\ref{fig:mpc_response}, this comparison allows the horizon and terminal weight to vary.
        Within this set, HEK-Control meets the 95\% CSTR criterion down to 0.4 times the nominal bound, compared with 0.8 for both baselines.
        Van der Pol improves relative to DMDc but not to Jacobian control.
        At the 0.4 CSTR bound, MPC with $H=20$ and $Q_z$ settles 98--100\% across ten data seeds, versus 22--26\% for DMDc and 22\% for Jacobian.
        The $H=5$ configuration at the 0.6 bound ranges from 60\% to 99\%, indicating greater sensitivity to the identification data.

        \begin{table}[t]
            \centering\footnotesize
            \captionsetup{font=footnotesize}
            \caption{Highest settling rates (\%) over twelve designs on an unencrypted $11\times11$ grid. Bounds are relative to nominal. Bold: at least 95\%.}
            \label{tab:scope}
            \setlength{\tabcolsep}{3pt}
            \begin{tabularx}{\columnwidth}{@{}Xcrrr@{}}\toprule
                Plant & Bound & Jacobian & DMDc & HEK-Control\\\midrule
                CSTR & 0.8 & \textbf{99} & \textbf{99} & \textbf{100}\\
                CSTR & 0.6 & 53 & 53 & \textbf{100}\\
                CSTR & 0.5 & 37 & 37 & \textbf{100}\\
                CSTR & 0.4 & 22 & 21 & \textbf{99}\\
                van der Pol & 0.6 & \textbf{100} & 1 & \textbf{100}\\
                van der Pol & 0.5 & \textbf{100} & 1 & 4\\\bottomrule
            \end{tabularx}
        \end{table}

        Terminal weighting also matters: with $H=5$ and $Q_z$, nominal-bound costs are comparable for Jacobian and HEK-Control MPC (19.04 and 19.34).
        At the 60\% bound, this HEK-Control design achieves 99\% grid settling over 8 min but does not settle from the nominal state within 4 min. Beyond the training region, the benefit can reverse: over twice the van der Pol region, the best tested HEK-Control and Jacobian designs settle 64\% and 82\%, respectively.

    \subsection{Computational Cost}
        Without SIMD, Koopman observable evaluation takes less than 0.002\% of the LQR processing time, which is dominated by encryption and homomorphic evaluation.
        Table~\ref{tab:timing} reports SIMD speedups and packed processing time relative to the sampling period.

        \begin{table}[t]
            \centering\footnotesize
            \captionsetup{font=footnotesize}
            \caption{SIMD results (30-run medians, Section~\ref{sec:settings}). Speedup is unpacked/packed processing time. $t_{\mathrm{pack}}/T_s$ is packed encryption, homomorphic evaluation, and decryption time divided by the sampling period. Initialization and network transport are excluded.}
            \label{tab:timing}
            \setlength{\tabcolsep}{3pt}
            \begin{tabularx}{\columnwidth}{@{}Xrr@{}}\toprule
                Control / plant & Speedup & $t_{\mathrm{pack}}/T_s$\\\midrule
                LQR / van der Pol & $6.2\times$ & 0.47\\
                LQR / reaction--diffusion & $32.6\times$ & 0.53\\
                LQR / DC motor & $25.0\times$ & 1.16\\
                CSTR MPC, $H=5$ & $6.2\times$ & 0.19\\
                CSTR MPC, $H=20$ & $56.5\times$ & 0.26\\\bottomrule
            \end{tabularx}
        \end{table}

        Packing reduces the motor uplink from 102 ciphertexts to one.
        For CSTR MPC with $H=20$ and $L=15$, uploaded/returned ciphertext counts decrease from 305/320 to 16/16, while 16 round trips remain.
        The maximum normalized-input-sequence error in this standalone test is $1.6\times10^{-6}$.

\section{Discussion}\label{sec:discussion}
    The results show how Koopman lifting connects nonlinear plants to encrypted linear computation.
    This section discusses what the control results and timing measurements imply for cloud deployment.

    \subsection{Cloud Control}
        Numerical agreement supports separating client-side controller preparation from encrypted server-side execution.
        Koopman lifting lets different nonlinear plants use the same form of encrypted matrix--vector operation.
        The CSTR results in Table~\ref{tab:scope} show that this connection can meet control objectives under tighter input bounds within the tested design set.
        The horizon, terminal-weight, and data sensitivity show that this benefit depends on the predictor and controller together, rather than encryption itself.

        Identification and controller design remain at the client.
        Extending outsourcing to these stages would require assessment of their confidentiality and of how coefficient errors affect subsequent closed-loop performance.
        The present numerical-error results concern only online control evaluation.

    \subsection{Deployment Requirements}
        Encryption and homomorphic evaluation dominate the measured cost, so packing addresses a larger workload than observable evaluation.
        The closed-loop simulations omit delays, although the corresponding unpacked computation times exceed the sampling period in the cases listed in Table~\ref{tab:timing}.
        A SIMD speedup does not by itself establish real-time feasibility: packed motor LQR still exceeds the sampling period.
        Packed CSTR MPC computation fits within the period in the standalone tests, but its 16 round trips must share the remaining time with other unmeasured processing.

        These results quantify the overhead of confidential cloud execution, not a speedup over plaintext control at the client.
        Shared resources and centralized deployment motivate outsourcing, while SIMD reduces its cryptographic overhead.
        Separate client and server hardware must be evaluated under the combined computation and communication delay before drawing conclusions about networked closed-loop performance.

\section{Conclusion}\label{sec:conclusion}
    This paper proposed Homomorphically Encrypted Koopman-Based Control (HEK-Control) for outsourcing control computations for nonlinear plants to a third-party cloud server without disclosing measurements or controller coefficients.
    A trusted plant-side client evaluates nonlinear state observables through Koopman lifting and uses a linear predictor identified in these coordinates for controller design.
    Identification and design are completed at the client, while the server applies existing encrypted linear operations to the encrypted coefficients and observables.

    LQR-based state feedback and input-constrained MPC served as representative applications.
    Closed-loop simulations using CKKS closely reproduced the corresponding unencrypted computations.
    Additional unencrypted CSTR evaluations met the settling criterion under tighter input constraints than the original-state comparison methods within a common set of controller designs.
    Performance on other plants can improve, remain comparable, or deteriorate with the design conditions.
    In particular, the van der Pol tests include conditions with lower settling rates than Jacobian control.
    The results support the use of encrypted linear computation for nonlinear plants, with the predictor and controller validated together over the intended operating region.

    Separate computation tests showed that SIMD packing reduces processing time and communication volume.
    Iterative MPC retains multiple communication round trips, so real-time deployment requires joint assessment of computation and communication.
    Future work will use separate client and server hardware with physical plants to evaluate delays, measurement noise, and closed-loop performance.
    Reducing the client workload and extending the framework to observer-based linear controllers are further directions.

\raggedend
\printreferences{references}

\begin{biography}
    \profile[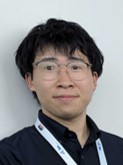]{Uta Kawakami}{%
    Uta Kawakami is currently a doctoral student at the Graduate School of Engineering, The University of Osaka, Japan.
    He graduated from The University of Electro-Communications, Japan, in 2024 and received the M.E. degree from the same university in 2026.
    His research interests include control engineering, autonomous driving, encrypted control, and robotics.
    He is a member of the IEEE and the IEEE Control Systems Society.
    }
    \profile[figures/sawadakenji2021_lab]{Kenji Sawada}{%
    Kenji Sawada received the Ph.
    D. degree in engineering from Osaka University, Japan, in 2009.
    He is currently a Professor in the Department of Mechanical Engineering, The University of Osaka, Japan.
    He joined The University of Electro-Communications as an Assistant Professor in 2009 and became an Associate Professor there in 2015, before taking up his current position in 2025.
    His research interests include robust control, control system security, human-in-the-loop systems, and intelligent transportation systems.
    He is a member of SICE, ISCIE, IEICE, JSME and IEEE.
    }
\end{biography}

\end{document}